\documentclass[a4paper,10pt,twocolumn]{article}
\usepackage{amssymb}
\usepackage{graphicx}
\usepackage{epsfig}
\usepackage{caption}

\begin{document}

\title{Boundary Dissipation in a Driven Hard Disk System}
\author{ P.L. Garrido$^{(1)}$ and G. Gallavotti$^{(2)}$\\ \\
$^{(1)}$ Institute `Carlos I' for Theoretical and Computational
Physics, \\
and Departamento de Electromagnetismo y F\'{\i}sica de la Materia,
\\University of Granada, 18071--Granada, Spain.
\\ \\
$^{(2)}$ Dipartimento di Fisica. INFN. \\
Universit{\`a} di Roma
"La Sapienza", 00185 Roma, Italy}
\date{\today}
\maketitle

\begin{abstract}
{\it A simulation is performed aiming at checking the existence of a well
defined stationary state for a two dimensional system of driven hard
disks when energy dissipation takes place at the system boundaries and
no bulk impurities are present.}

\end{abstract}
\vglue 0.5cm

Bulk dissipation is often assumed to explain stationary
states in driven systems as in the well known example of Drude's
theory of electrical conduction where three mechanisms act over a
given interacting particle system:
\\
(1) a constant force that accelerates the particles in a given
direction,
\\
(2) a thermal bath that should drive the system to an equilibrium state
absorbing energy excess due
to the action of the driving and
\\
(3) an array of bulk impurities that introduce a strong chaotic behavior on the
particle dynamics.

The stationary state is characterized by a net current of particles in
the field direction generated by the external work per unit time
done by the field over the particles equals the heat flux absorbed by
the thermal bath. The existence of such stationary state is physically
intuitive: bulk forcing is compensated by bulk dissipation and it can
be seen in several computer simulations (see for instance
\cite{Yuge}). Moreover, it is expected that the thermostat
model used does not influence the system statistical properties
\cite{AES01,Gallavotti}, but some other dynamical properties may depend on
the particular dissipation scheme used \cite{RKlages}.
\begin{figure}[ht]
\begin{center}
\includegraphics[width=6.cm]{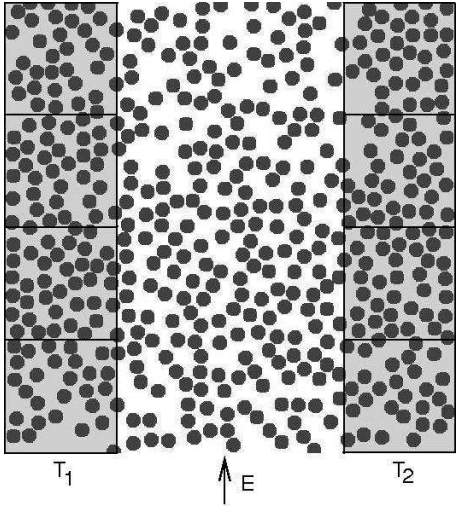}
\end{center}
\caption{\it Typical configuration of the model simulated. The disks in the
white part (bulk
particles) are accelerated in the $y$ direction by a driving field $E$. The
disks at the grey boxes
act as thermal baths, {\and} keep constant their overall respective kinetic
energies (temperatures
$T_1$ and $T_2$) for all times. The disks interact each other with normal
elastic collisions. The
center of the disks may also elastically collide with the walls (black lines).}
\label{figure1}
\end{figure}

Existence of a stationary state is, however, not so clear if the
action of the thermostat is at the system boundaries and no impurities
are present: the field tends to align the particles
trajectories and the boundaries introduce a disorder ``transversal''
to the field and this is a bulk versus a surface effect.

In this case, a similar system studied in hydrodynamics seems to lead
to a well defined stationary state: a model for the Poiseuille flow in
the weak-flow regime. There, a group of interacting particles are
subject to a small external gravitational field that drives the
particle flow between two parallel plates that are kept at constant
temperature while strongly interacting via long range forces. Several
computer simulations of these system confirm that the
heat generated by a bulk force is efficiently removed by the
thermostats even though the dynamics in the bulk of the system is
conservative: and the system evolves tending to a stationary state
\cite{TOD,AES01}. This is different from other studies showing
thermostats efficiency in cases in which thermostats act on the system
through its boundaries but the driving force also acts on the boundary
only, \cite{BCL98}.

It appears uncertain, more generally, whether the boundaries and the
likely intrinsic chaotic behavior of the system would compensate the
action of a strong external field leading to establish a stationary
state (the question has been recently again raised in the literature,
\cite{Gallavotti,Ga06,Ru06}, and called the problem of {\it
efficiency} of a thermostat mechanism) we consider worth studying in
other cases the problem of whether a thermostat acting only on the
boundary of the system is efficient enough to thermalize a system
subject to {\it bulk} driving forces looking for other similar
instances to add to the basic result in \cite{TOD,AES01}. The latter
has been, to our knowledge, the first to show that such thermostats
can actually be efficient to remove the heat generated by a bulk force
even though the dynamics in the bulk of the system is conservative:
and the system evolves tending to a stationary state.

A thermostat is ``efficient'' if it absorbs enough energy
(``thermalizes'') to forbid indefinite energy growth of a forced
system.

In this note therefore we consider a system of hard disks under the
action of a driving field and check that, within the range of external
force strength that we are able to simulate, it appears to reach a
well defined stationary state. Even for ``strong'' driving fields, although
the thermostat acts only near the system boundaries, {\it i.e.}
through very short range forces (in fact we consider hard core forces) between
pairs of particles of the system and of the thermostats and between
``mixed pairs'' of system and thermostats particles.

The model consists of hard disks confined to a {\it unit box} and
initially placed on a triangular lattice structure. The disks radius,
$r$, is fixed so that the maximum number $N_{max}$, of particles in
close-packing for a unit surface have a preassigned value. That is,
$r=\left(\rho_{cp}/N_{max}\pi\right)^{1/2}$ where
$\rho_{cp}=\pi/2\sqrt{3}\sim .9069$ is the close-packing mass density
for hard disks: we take here $N_{max}=10^4$ and hence $r=5.37
10^{-3}$.

\begin{figure}[h]
\begin{center}
\includegraphics[width=7.cm,angle=180]{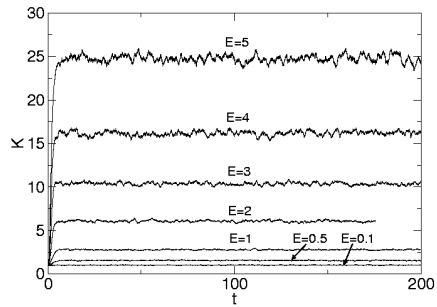}
\includegraphics[width=7.cm,angle=180]{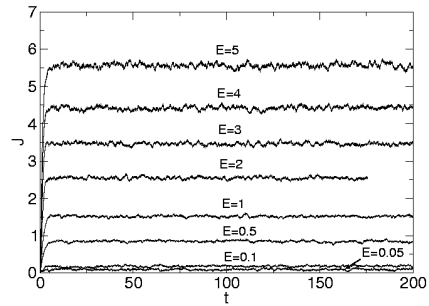}
\end{center}
\caption{\it Evolution of $K=\frac1{2N}\sum_{i=1}^{N_{bulk}}
  (v^2_{xi}+v^2_{yi})$, kinetic energy per particle (top) and particle current
$J=\frac1{N}\sum_{i=1}^{N_{bulk}}v_{yi}$
   on the $y$ direction (bottom) for the
bulk disks and for different values of the driving field.} \label{figure2}
\end{figure}

The box will be divided in three parts (see figure \ref{figure1}): a central
part of width
$1-2\alpha$ (``bulk'') with top and bottom identifies (``vertical periodic
boundary conditions'')
and two equal lateral parts of width $\alpha$ (``baths'').

The actual number of disks placed in the bulk and in the bath parts are
controlled by the
corresponding densities of disks: $\rho_{bulk}$ and $\rho_{bath}$. In our
simulations we have
chosen $\rho_{bulk}=0.4$ and $\rho_{bath}=0.5$. This implies that in our
simulations the number of
disks present in the bulk part is $N_{bulk}=2301$, at each bath $N_{bath}=1318$
and the total
number is then $N=4937$.

Disks dynamics depends on the sector they are. If a disk is in the bulk part it
is subject {\it
between collisions} to a uniform acceleration of magnitude $E$ in the $y$
direction while in the
$x$ direction its velocity keeps constant. The disks pertaining to the baths
move at constant speed
along its velocity vector.

In all cases, when the boundaries of two disks meet (``collision'')
they undergo an elastic collision. When the {\it center} of a disk
hits any of the walls that define the region in which it is contained,
it is elastically reflected. In this way we manage to keep disks
confined at their respective regions and particles from the bulk may
interact with particles in the thermal baths only across the walls.

Four equidistant walls along the $x$ direction, see figure
(\ref{figure1}), in each bath prevent the disks of the bath having a
net movement along the $y$ direction induced by the interaction with
the disks on the bulk.

The disks in each bath keep their total kinetic energy constant:
$K_{1,2}=N_{bath}T_{1,2}$.  This is achieved by the following
prescription: when a disk from the bath collides with another of the
bulk, the increment, $\Delta$, of kinetic energy (positive or
negative) that the bath particle suffers is immediately shared with
the other particles of the bath by rescaling their speeds by the
factor $(1+\Delta/K_{1,2})^{-1/2}$ respectively.

Initially we let the system evolve during $100\,N$ collisions with
$E=0$: this is empirically sufficient to homogenize it spatially; next
we turn on the driving field. Then, we take measures at intervals of
$N$ collisions during $10^5\,N$ collisions. We have simulated the
cases with driving fields $E=0.0001$, $0.0005$, $0.001$, $0.005$,
$0.01$, $0.05$, $0.1$, $0.5$, $1$, $2$, $3$, $4$ and $5$ with
$T_1=T_2=1$.

Figure \ref{figure2} shows typical evolutions of the kinetic energy per
particle and the average
current along the field direction of the bulk disks. After some short initial
transient, apparently
the system reaches a stationary state with a well defined current and kinetic
energy.

\begin{figure}
\begin{center}
\includegraphics[width=6.cm,angle=-90]{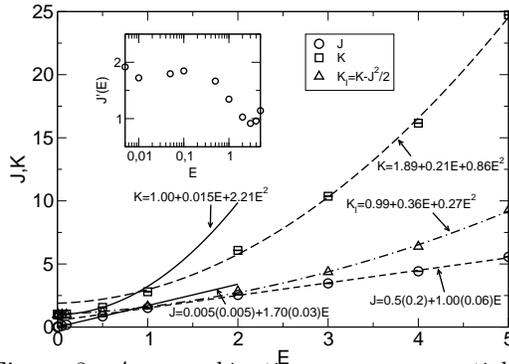}
\end{center}

\vglue-1.2truecm
\caption{\it Average kinetic energy per particle ($K$) and averaged particle
current ($J$) at the
system stationary state for the bulk disks. Lines are linear or square
regression fits of the data.
$K_I$ is the internal kinetic energy per particle. Inset shows the forward
discrete derivative of
$J(E)$. Error bars are present.} \label{figure3}
\end{figure}

Figure \ref{figure3} shows the measured stationary values of the kinetic energy
per particle and
the particle current. We see that the hard disk system follow a
nonlinear current-field response with: $J\simeq E$ and
$K\propto E^2$ for large $E$ while for small fields ($E<1$)
we see Ohm's law with a
{\it larger} conductivity: $J\simeq 1.7 E$ (see inset in Figure
\ref{figure3}).

This behavior is consistent with the picture that the driving (that tends to
align particles) seems
to generate an intense enough interaction with the boundaries that disorders
efficiently the bulk
particles. If we consider the internal kinetic energy, $K_I=K-J^2/2$, (total
kinetic energy minus
the kinetic energy of the center of mass in the vertical direction) then we
find that $K_I$
increases quadratically with $E$ exhibiting such disordering effect of the
boundary interactions
which has the effect of lowering the conductivity.

\begin{figure}[h]
\begin{center}
\includegraphics[width=5.cm,angle=-90]{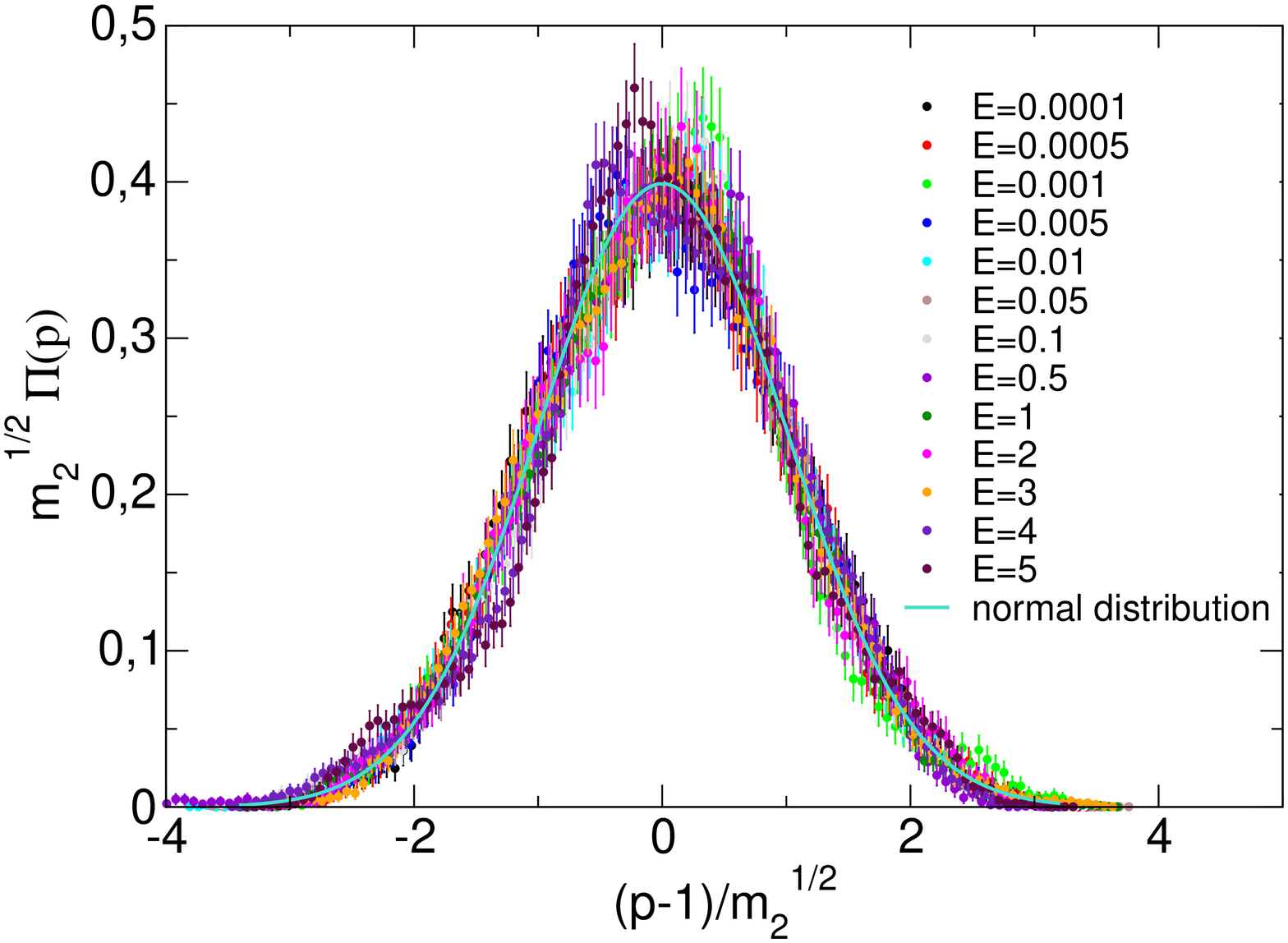}
\includegraphics[width=5.cm,angle=-90]{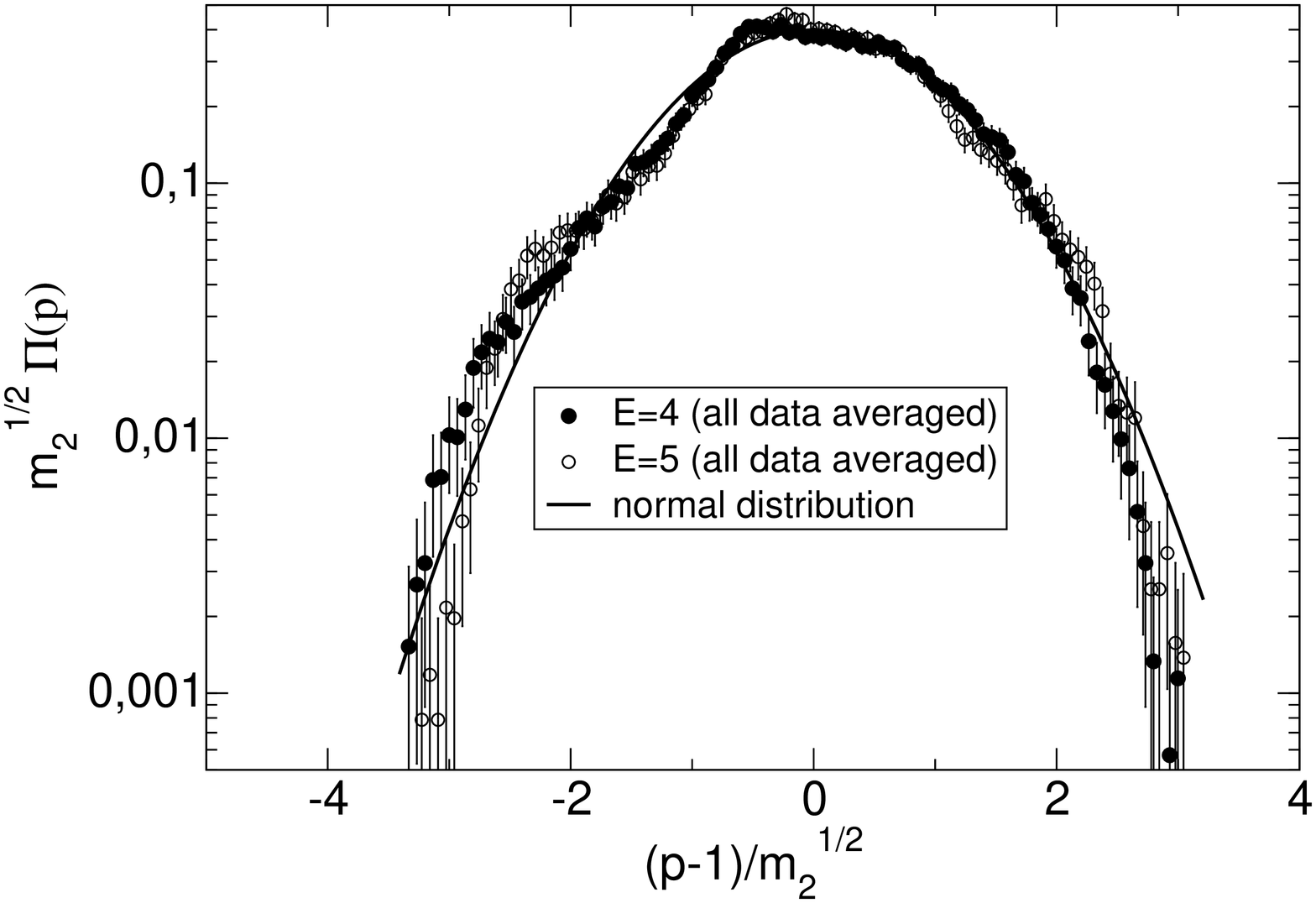}
\end{center}
\caption{\it Top: The measured probability distribution of $p(t)=J(t)/J$,
$\Pi(p)$, at the
stationary state for different electric fields. $m_2(E)$ is the observed
variance of the $p(t)$
variable. Solid line is the normal distribution with zero average and variance
one. Bottom:
$\ln\Pi(p)$ vs. $(p-1)/m_2(E)$ plot for $E=4$, $E=5$.} \label{figure4}
\end{figure}

\begin{figure}
\begin{center}
\includegraphics[width=6.cm,angle=-90]{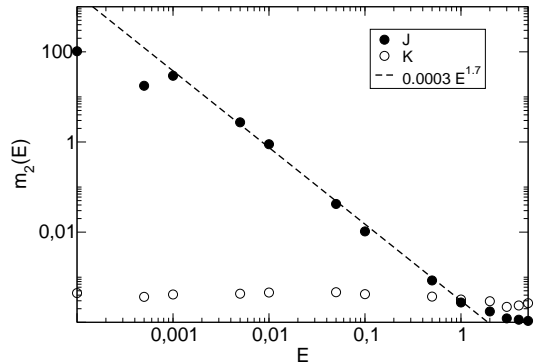}
\end{center}
\caption{\it The observed variance $m_2(E)$ of $J(t)/J$ and $K(t)/K$
versus the electric field.  Dashed line is guide to the eye showing
the power law behavior $E^{-1.7}$.}
\label{figure5}
\end{figure}

We have also studied the fluctuations distribution of $J(t)$ and
 $K(t)$ around
its stationary
average value. In particular the top of figure \ref{figure4} shows the
distribution of the observed
values of $p(t)=J(t)/J$, $\Pi(p)$. We see that for $E\leq 1$ the measured
distribution is
compatible with a gaussian distribution with an average value $1$ and a
variance $m_2(E)$ that
depends on the electric field. However, systematic deviations from gaussianity
is observed for
large electric field (see bottom of figure \ref{figure4}). Moreover, in figure
\ref{figure5} we
show the variance, $m_2(E)$, of $J(t)/J$ and $K(t)/K$ for different values of
$E$. We see while
$m_2(E)$ for the kinetic energy depends weakly on the electric field, the
variance for the current
decays with $E$ as a power: $m_2(E)\simeq E^{-1.7}$ from $E=0.001$ to $E=1$.
That is, for $E<<1$ a
large set of particles are able to move in the $-E$ direction. This behavior is
strongly suppressed
as the field increases and, for $E>1$ most of the particles move along the
field. Note that the
large error bars due to the limited amount of data obtained in this simulation
obscures the
analysis of the large deviations properties of $p(t)$ and so we are not able to
check the
fluctuation theorem as proposed in \cite{Ga06}.

Finally we have computed the correlation between the current and the energy at
the stationary
state: $C_{J-K}=\langle J(t) K(t) \rangle/JK -1$. For all cases $C_{J-K}$ is
compatible with zero
within error bars. That is, on the stationary state both magnitudes are
uncorrelated and, for
instance, the value of the system kinetic energy is independent on the sign of
the current for a
given configuration.

Further investigations could be done by changing the temperatures of the
thermostats to two
different values.

\vfill\eject

\end{document}